\documentclass[a4paper,12pt]{article}
\usepackage{orcidlink}
\usepackage{amsmath,amssymb}
\usepackage{comment}
\usepackage{bm}
\usepackage{graphicx}
\usepackage[dvipdfmx]{epsfig}
\usepackage{epsfig}
\usepackage{braket}
\usepackage{color}
\usepackage{caption}
\usepackage{tikz}
\usepackage[compat=1.1.0]{tikz-feynman}
\usepackage{enumitem}
\usepackage{siunitx}
\usepackage{hyperref}
\hypersetup{colorlinks,allcolors=black}
\usepackage[nosort,noadjust]{cite}

\usepackage[normalem]{ulem}
\usepackage{xcolor}

\allowdisplaybreaks

\DeclareMathAlphabet{\pazocal}{OMS}{zplm}{m}{n}

\usepackage{letltxmacro}
    \LetLtxMacro\oldl\l
    \DeclareRobustCommand{\l}{\ifmmode\lambda\else\oldl\fi}
    \LetLtxMacro\oldpr\partial
    \DeclareRobustCommand{\pr}{\ifmmode\partial\else\oldpr\fi}
    \LetLtxMacro\oldd\dagger
    \DeclareRobustCommand{\d}{\ifmmode\dagger\else\oldd\fi}
    \LetLtxMacro\oldno\noindent
    \DeclareRobustCommand{\no}{\ifmmode\noindent\else\oldno\fi}

\begin{document}
\title{
\begin{flushright}
\ \\*[-80pt] 
\begin{minipage}{0.2\linewidth}
\normalsize
OU-HET-1288 \\*[50pt]
\end{minipage}
\end{flushright}
{\Large \bf 
Vacuum Structure of an Extended \\ Standard Model with $U(1)_D$ Symmetry  
\\*[20pt]}}
\author{ 
\centerline{
Apriadi Salim Adam\orcidlink{0000-0001-6587-5156}$^{1}$\footnote{\texttt{E-mail address: apriadi.salim.adam@brin.go.id}}, 
Yunita Kristanti Andriani\orcidlink{0000-0002-8980-2274}$^{2,4}$\footnote{\texttt{E-mail address: ykandriani2871@gmail.com}},
~and Bayu Dirgantara\orcidlink{0000-0002-3308-3055}$^{3}$\footnote{\texttt{E-mail address: bayudirgantara@unhas.ac.id}}
}
\\*[20pt]
\centerline{
\begin{minipage}{\linewidth}
\begin{center}
$^1${\it \normalsize
Research Center for Quantum Physics, National Research and Innovation Agency (BRIN), 
South Tangerang 15314, Indonesia} \\*[5pt]
$^2${\it \normalsize
Department of Physics, Faculty of Mathematics and Natural Sciences, Universitas Negeri Malang, Jl. Semarang 5, Malang 65145, Indonesia} \\*[5pt]
$^3${\it \normalsize
Theoretical and Computational Physics Laboratory, Department of Physics, Hasanuddin University, Makassar, South Sulawesi 90245, Indonesia}
\\*[5pt]
$^4${\it \normalsize Department of Physics, The University of Osaka, Toyonaka, Osaka 560-0043, Japan}
\end{center}
\end{minipage}}
\\*[50pt]}
\date{
\centerline{\small \bf Abstract}
\begin{minipage}{0.9\linewidth}
\medskip 
\medskip 
\small
In this research, we investigate the vacuum structure of an extended standard model with a $U(1)_D$ global symmetry. The scalar sector consists of two $SU(2)$ doublets as well as one complex singlet and one real singlet, resulting in a more complicated vacuum structure compared to that of the Standard Model. We analyze various theoretical constraints, including the conditions for being bounded from below, the existence of a global minimum, and perturbativity up to the Planck scale. Additionally, we consider experimental constraints from the Higgs invisible decay. Through a detailed statistical analysis using numerical methods, we show that the extended scalar potential can accommodate a stable vacuum while satisfying both theoretical and experimental constraints for a small region of the parameter space.
\end{minipage}
}
\begin{titlepage}
\maketitle 
\thispagestyle{empty}
\end{titlepage}
\section{Introduction}

The 2012 discovery of the Higgs boson \cite{ATLAS:2012yve,CMS:2012qbp} at the Large Hadron Collider (LHC) marked a major milestone for the Standard Model (SM), confirming its predictions in both the hadron and leptonic sectors. Still, the question of vacuum stability remains open due to the large top quark Yukawa coupling ($y_t\sim \mathcal{O}(1)$) and the measured mass of around 173 GeV \cite{ParticleDataGroup:2022pth}. The Higgs potential starts to be unstable at high energies and turns negative at a scale of around $10^{10}$ GeV \cite{Isidori:2001bm,Degrassi:2012ry}. This implies the need for a new theory beyond the Standard Model (BSM).

Various BSM theories have been proposed to address vacuum instability, often incorporating new symmetries (either global or local) \cite{Mohapatra:1974hk,Mohapatra:1974gc} or/and new scalar fields as an extension.
One popular approach involves extending the SM by adding an extra $U(1)$ symmetry \cite{Ma:2002tc,Montero:2007cd,Adhikari:2015woo}, which introduces new gauge bosons and scalar fields. For instance, models such as the $U(1)_{B-L}$ \cite{Davidson:1978pm,Marshak:1979fm,Foot:1989ts,Dulaney:2010dj,FileviezPerez:2010gw,FileviezPerez:2013eoz,Ho:2016aye} have been extensively studied due to their ability to explain the smallness of neutrino masses via the seesaw mechanism and to predict new $Z'$ gauge bosons that can be searched for in collider experiments. Similarly, $U(1)_{L_\mu - L_\tau}$ symmetry models \cite{Ma:2001md,Baek:2008nz,Altmannshofer:2014cfa}, which involve the difference between muon and tau lepton numbers, have been considered to address anomalies in muon $g-2$ measurements and to provide dark matter candidates. For the vacuum investigation of various $U(1)$ extended models, kindly see Refs.\cite{Sher:1988mj,Coriano:2014mpa,Das:2019pua,Dirgantara:2023vsi}.

In this work, we focus on an extended model with additional $U(1)_D$ global symmetry \cite{Dutta:2022knf} and present an analysis of its vacuum stability. The model comprises an SM-like scalar field, one additional SU(2) scalar doublet, one real scalar singlet, and one complex scalar singlet.
The two singlets acquire vacuum expectation values (VEVs) that break the additional symmetry, while the doublet $\eta$ is assumed as an inert doublet with zero VEV, and facilitates dark sector dynamics via a $Z_2$ symmetry. These extra scalars lead to a richer vacuum structure and mixing patterns in the CP-even, CP-odd, and charged scalar sectors. 

The mixing between the SM-like Higgs and a light scalar from the singlet sector enables interaction with dark fermions, making the model testable through Higgs invisible decay and direct detection of self-interacting dark matter via scattering \cite{Dutta:2022knf}.
At the same time, ensuring perturbativity and boundedness from below in all directions of field space at large field values remains a crucial consistency check. We study the vacuum structure of the model in detail by analyzing theoretical conditions and exploring the parameter space numerically while adopting systematic approaches from Refs.\cite{BhupalDev:2018xya,Chauhan:2019fji,Frank:2021ekj}.
We also investigate the scalar mass spectrum and include the Higgs invisible decay constraint to generate the allowed parameter space.

The structure of the paper is as follows: Section \ref{sec2} presents the $U(1)_D$ extended model, focusing on the scalar potential and the resulting scalar mass spectrum. In Section \ref{sec3}, we address the minimum of the scalar potential and evaluate its vacuum stability. We present the copositivity conditions necessary for a bounded potential and identify regions with a true electroweak vacuum through numerical minimization. Section \ref{sec4} analyzes the renormalization group evolution of couplings and tests the perturbativity up to the Planck scale. In Section \ref{sec5}, we investigate the viable parameter space considering the obtained theoretical and experimental constraints, particularly from Higgs invisible decay. Section \ref{sec6} is devoted to our summary and an outlook. In Appendix \ref{app.A}, we introduce the complete sets for eight different cases related to the copositivity conditions derived from the quartic coupling matrix. Appendix \ref{app.B} provides the beta function equations at the one-loop level for the gauge and the Yukawa couplings.

\section{The $U(1)_{D}$ extended model}
\label{sec2}
\subsection{The scalar potential} \label{sec.2.1}
We consider the SM with an extended $U(1)_D$ global symmetry \cite{Dutta:2022knf}. The SM Higgs doublet, $H$, is not charged under this global symmetry. In the fermion sector, a vector-like Dirac fermion $\chi$ and three heavy right-handed neutrinos $\nu_{R_i}$ are introduced with their quantum number assignments under $SU(2)_L \times U(1)_Y \times U(1)_D$ gauge group as follows
\begin{equation}
    \chi : \left( 1,0,\frac{1}{2} \right), \qquad \nu_{R_i} : \left(1,0,-1\right),
\end{equation}
where $i=1,2,3$. 
The scalar field content is written in irreducible representations 
as{\footnote{In Ref.\cite{Dutta:2022knf}, the original model introduces heavy doublets $X_i\,(i=1,2)$ that do not engage in the low-energy process and have only implications in the leptogenesis scenario and Dirac-type neutrino mass generation. The corresponding terms of the $X$ interaction are $y_{\nu_{R}} \overline{L}\tilde{X}\nu_R$ and $\rho \Phi'^\ast X^\dagger H$. These terms lead to a dimension-five operator after integrating out at the low-energy, $\mathcal{O}_5 = y\frac{\rho}{M^2_X} \overline{L} H \Phi' v_R$. Therefore, in this work, we only focus on the effective potential of the model (Eq.\eqref{eq.1}).}}
\begin{align}
     {H: \left(2,1,0\right),} \qquad \eta : \left( 2,1,\frac{1}{2} \right), \qquad \Phi^\prime : \left( 1,0,1 \right), \qquad \Phi : \left( 1,0,0 \right).
\end{align}
As presented above, $\eta$ is the scalar doublet, while $\Phi$ and $\Phi'$ are the singlet scalars. 
On the other hand, the scalar $\eta$ is assumed to have zero VEV to preserve the $\mathbb{Z}_2$ symmetry\footnote{$\eta$ plays a role in facilitating the transfer of the lepton asymmetry from the dark to the visible sector. The detailed analysis of the leptogenesis scenario is presented in Ref.\cite{Dutta:2022knf}.}. Under the $\mathbb{Z}_2$ symmetry, the scalar $\eta$ is CP-odd, while the others are CP-even. These additional scalar fields carry nontrivial charge under $U(1)_D$ symmetry, except the scalar $\Phi$.

The scalar potential of the proposed model under the imposed symmetry is given by,
\begin{equation} \label{eq.1}
    \begin{split}
        V &= \mu^2_1 (\eta^\d \eta) - \mu^2_2 H^\d H + \frac{1}{2} \mu^2_3 \Phi^2 + \frac{1}{3} \mu_4 \Phi^3 - \mu^2_5 (\Phi'^\d \Phi') + \frac{\mu_6}{\sqrt{2}} \Phi (H^\d H) \\
        & \quad + \frac{\mu_7}{\sqrt{2}} \Phi (\Phi'^\d \Phi') + \l_1 (\eta^\d \eta)^2 + \l_2 (\eta^\d \eta)(H^\d H) + [\l_3 (\eta^\d H)^2 + \mathrm{H.c.}] \\
        & \quad + \l_4 (H^\d H)^2 + \frac{1}{4} \l_5 \Phi^4  + \l_6 (\Phi'^\d \Phi')^2 + \frac{\l_7}{2} H^\d H \Phi^2 + \l_{8} H^\d H (\Phi'^\d \Phi') \\
        & \quad + \frac{\l_9}{2} \Phi^2(\Phi'^\d \Phi') + \l_{10} (\eta^\d H)(H^\d \eta) + \l_{11} (\eta^\d \eta)(\Phi'^\d \Phi') + \l_{12} \eta^\d \eta\Phi^2,
    \end{split}
\end{equation}
where the couplings $\mu_{i}$ have a mass dimension and represent the soft-breaking terms while the other couplings are dimensionless. Note that the original model in \cite{Dutta:2022knf} does not include the quartic terms $\l_{10} (\eta^\d H)(H^\d \eta)$, $\l_{11} (\eta^\d \eta)(\Phi'^\d \Phi')$, and $\l_{12} (\eta^\d \eta)\Phi^2$. For our purpose of this study, we add these terms in our analysis. 
The parameterization of the scalar fields, in which $\Phi$ is considered to be a real scalar, is
\begin{align}
    \Phi' = \frac{1}{\sqrt{2}} \left(v_1 + h_1 + i \phi_1\right),
    \qquad H = \begin{pmatrix}
            H^+\\
            \frac{1}{\sqrt{2}} \left(v_2 + h_2 + i\phi_2 \right)	
    \end{pmatrix}, \\
    \Phi = v_3 + h_3,
    \qquad \eta = \begin{pmatrix}
            \eta^+ \\
            \frac{1}{\sqrt{2}} \left(v_4 + h_4 + i\phi_4\right)	
    \end{pmatrix}.
\end{align}
After the symmetry breaking takes place, the scalar fields attain their VEVs as
\begin{align} \label{eq.2}
    \langle\Phi'\rangle = \frac{v_1}{\sqrt{2}},
    && \langle H\rangle = \frac{1}{\sqrt{2}} \begin{pmatrix}
            0\\
            v_2	
    \end{pmatrix},
    && \langle\Phi\rangle = v_3,
\end{align}
while the VEV of the scalar field $\eta$ is set to be zero, $v_{4}=0$. The VEVs are also assumed to satisfy the following hierarchy:
\[v_1 > v_2 \gg v_3,\]
in which this assumption agrees with the analysis of the relevant phenomenology done in Ref. \cite{Dutta:2022knf}.

Due to the non-zero VEVs, the chosen parameters $\mu_{i}$ ($i=1,..,7$) and $\lambda_{j}$ ($j=1,..,12$) in the potential \eqref{eq.1} will give rise to a vacuum configuration for certain scalars.
The minimization conditions of the potential related to the assigned VEVs are computed as the first-order derivatives of the scalar potential, $(\pr{V}/\pr{v_i})=0$ $(i=1,2,3)$, and we obtain
\begin{align}
    \frac{\pr{V}}{\pr{v_1}} &= v_1 \left( - \mu^2_{5} + \l_{6} v^2_1 + \frac{\mu_7}{\sqrt{2}} v_3 + \frac{\l_8}{2} v^2_2 + \frac{\l_9}{2} v^2_3 \right) = 0, \label{eq4} \\
    \frac{\pr{V}}{\pr{v_2}} &= v_2 \left( - \mu^2_2 + \l_4 v^2_2 + \frac{\mu_6}{\sqrt{2}} v_3 + \frac{\l_7}{2} v^2_3 + \frac{\l_{8}}{2} v^2_1\right) = 0, \label{eq5} \\
    \frac{\pr{V}}{\pr{v_3}} &= v_3 \left( \mu^2_3 + {\mu_4 v_3} + {\l_5 v^2_3} + \frac{\l_7}{2} v^2_2 + \frac{\l_{9}}{2} v^2_1 \right) + \frac{\mu_6}{2 \sqrt{2}} v^2_2 + \frac{\mu_7}{2 \sqrt{2}} v^2_1 = 0. \label{eq6}
\end{align}
Corresponding to these minimum, the scalar mass spectrum and its mixing matrices will be presented in the following subsection.

\subsection{Scalar mass spectrum} \label{sec.2.2}

The potential in  Eq.\eqref{eq.1} has a rich scalar mass spectrum, encompassing CP-even, CP-odd, and charged scalar fields. These scalars arise as a consequence of spontaneous symmetry breaking and mixing among the scalar fields. In this subsection, we present the scalar mass spectrum, highlighting the physical states and their corresponding mass eigenvalues.

\subsubsection{Decoupled scalar}

In this model, the scalar $\eta$ has decoupled from the other scalar fields at a high-energy level; thus, one can directly obtain its squared effective mass $\bar{m}^2_{h_4}$ that can be extracted from the potential (Eq.\eqref{eq.1}). It is explicitly given by
\begin{equation} \label{eq.mass}
    \bar{m}_{h_4}^2 = \mu_1^2 + \frac{1}{2} \left(\lambda_2 + 2\lambda_3 +\lambda_{10}\right) v_2^2 + \frac{1}{2} \lambda_{11} v_1^2 + \lambda_{12} v_3^2.
\end{equation}
Since $\eta$ does not mix with other scalar fields, its mass is independently determined and remains unaffected by the diagonalization process of the scalar mass matrix. One should note that the above equation is slightly different from the effective mass of the decoupled scalar derived in \cite{SalimAdam:2023lbm} due to the additional terms in the potential (Eq.\eqref{eq.1}). 

\subsubsection{Charged scalar}

In this model, we write down the squared mass matrix of the charged scalars in the basis $(H^+, \eta^+)$. After the scalar fields gain their VEVs, the squared mass matrix reads
\begin{equation}
    \mathcal{M}_\text{charged}^2 = \begin{pmatrix}
        - \frac{1}{2} (2\lambda_3 + \lambda_{10}) v_4^2 & \frac{1}{2} (2\lambda_3 + \lambda_{10}) v_2 v_4 \\
        \frac{1}{2} (2\lambda_3 + \lambda_{10}) v_2 v_4 & - \frac{1}{2} (2\lambda_3 + \lambda_{10}) v_2^2
    \end{pmatrix}.
\end{equation}
In fact, the above mass matrix has the rank of one. This implies that there is one massless charged Goldstone boson eaten by the charged SM gauge boson $W^\pm$. Recalling $v_4=0$, the remaining physically charged scalar mass is then given by
\begin{equation}
    m_{\tilde{\eta}}^2 = - \frac{1}{2} (2\lambda_3 + \lambda_{10}) v_2^2.
\end{equation}

\subsubsection{CP-odd scalar}
The squared mass matrix of the CP-odd scalar is written in the basis $(\phi_1,\phi_2,\phi_4)$. The expansion of the fields leads to the following squared mass matrix,
\begin{equation}
    \mathcal{M}_\mathrm{odd}^2 = \begin{pmatrix}
        0 & 0 & 0 \\
        0 & - 2 \lambda_3 v_4^2 & 2 \lambda_3 v_2 v_4 \\
        0 & 2 \lambda_3 v_2 v_4 & - 2 \lambda_3 v_2^2
    \end{pmatrix}
\end{equation}
Similar to the above singly-charged one, the rank of this squared mass matrix is one. Therefore, there are two massless fields. One of which is an unphysical neutral Goldstone boson $A_2$ absorbed by the neutral SM boson $Z$, while the other is a physical pseudo-scalar $A_1$. Recalling the previous assumption $v_4=0$ (see Sec.\ref{sec.2.1}), the remaining pseudo-scalar $A_3$ is massive with a mass $m_{A_3}^2 = - 2 \lambda_3 v_2^2$.

\subsubsection{CP-even scalar}

The remaining CP-even scalar fields are mixed, forming a $3 \times 3$ symmetric mass matrix in the $(h_1,h_2,h_3)$ basis.  This mass matrix is expressed as,
\begin{equation} \label{eq.26}
        \mathcal{M}^2_\mathrm{even} =
        \begin{pmatrix}
            2 \l_{6} v^2_1 & \l_{8} v_1 v_2 & \frac{\mu_7}{\sqrt{2}} v_1 + \l_{9} v_1 v_3 \\
            \l_{8} v_1 v_2 & 2 \l_4 v_2^2 & \frac{\mu_6}{\sqrt{2}}v_2 + \l_{7} v_3v_2\\
            \frac{\mu_7}{\sqrt{2}} v_1 + \l_{9} v_1 v_3 & \frac{\mu_6}{\sqrt{2}}v_2 + \l_{7} v_3v_2 & 2 \l_5 v_3^2 +\mu_4 v_3 - \frac{1}{2\sqrt{2} v_3} \left(\mu_6 v_2^2 + \mu_7 v_1^2\right)
       \end{pmatrix}.
\end{equation}
To obtain their physical mass on a mass basis, one must diagonalize it. This can be realized by introducing an orthogonal rotation matrix $\mathcal{O}$, such that
\begin{equation}
     \mathcal{O} \mathcal{M}_\mathrm{even}^2 \mathcal{O}^T = \text{diag}\left(m^2_{H_1},m^2_{H_2},m^2_{H_3}\right),
\end{equation}
where $m^2_{H_1}$, $m^2_{H_2}$, and $m^2_{H_3}$ are the eigenvalues of $\mathcal{M}^2_\mathrm{even}$ and represent the squared masses of the physical CP-even scalar states $H_1$, $H_2$, and $H_3$, respectively. By doing so, we can explicitly write
\begin{equation} \label{eq13}
    \begin{pmatrix}
        H_1 \\ H_2 \\ H_3
    \end{pmatrix} = \mathcal{O} \cdot
    \begin{pmatrix}
        h_1 \\ h_2 \\ h_3
    \end{pmatrix},
\end{equation}
and the form of the rotation matrix $\mathcal{O}$ is given by \cite{Ouazghour:2018mld}
\begin{equation} \label{eq14}
     \mathcal{O} = \begin{pmatrix}
        c_{\beta_1} c_{\beta_2} & s_{\beta_1} c_{\beta_2} & s_{\beta_2} \\
        -(c_{\beta_1} s_{\beta_2} s_{\beta_3} + s_{\beta_1} c_{\beta_3}) & c_{\beta_1} c_{\beta_3} - s_{\beta_1} s_{\beta_2} s_{\beta_3} & c_{\beta_2} s_{\beta_3} \\
        - c_{\beta_1} s_{\beta_2} c_{\beta_3} + s_{\beta_1} s_{\beta_3}) & -(c_{\beta_1} s_{\beta_3} + s_{\beta_1} s_{\beta_2} c_{\beta_3}) & c_{\beta_2} c_{\beta_3}
     \end{pmatrix},
\end{equation}
where the mixing angles $\beta_1$, $\beta_2$, and $\beta_3$ determine the degree of mixing among the CP-even scalars. These angles are constrained within the range:
\begin{equation}
     -\frac{\pi}{2} \leq \beta_i \leq \frac{\pi}{2}.
\end{equation}
By using the above procedure, we obtain numerically the eigenvalues of the squared mass matrix, i.e., $m_{H_i}^{2}$ ($i=1,2,3$).

\section{Minimization and vacuum stability} 
\label{sec3}
In this section, we explore the necessary conditions to achieve a good vacuum stability, focusing on conditions for a stable and physically viable vacuum. For this purpose, we apply copositivity criteria to ensure the boundedness of the potential and address the global minimum condition through a numerical analysis.
\subsection{Copositivity conditions} \label{sec3.2}
The quartic terms in the potential \eqref{eq.1} are the adequate terms to carry out the analysis of the vacuum stability. The other terms with dimensionful coupling, such as mass and soft-breaking terms, can be ignored due to a large limit value in the field configuration. We denote these quartic terms as potential $V^{(4)}$ and it is positive-definite, $V^{(4)} > 0$. This requirement is also known as the Bounded from Below (BfB) condition. 

To study the vacuum stability of a model, copositivity conditions are essential for evaluating its potential's boundedness \cite{Kannike:2012pe}. 
The potential with a biquadratic form of fields is said to be bounded from below if the coupling matrix is \textit{copositive} (conditionally positive). 
\begin{equation*}
    ax^2 + bx + c \geq 0,
\end{equation*}
with $x\in \mathbb{R}$, the copositivity condition requires
\begin{equation}
    a\geq0,\,\,c\geq0,\,\,b+2\sqrt{ac}\geq0.
\end{equation}
Since we are dealing with symmetric coupling matrices, it is necessary to introduce the term 'copositive matrix'. In other words, this copositive matrix has a similar definition to a positive-definite matrix. \textit{A symmetric matric $\bm{\Lambda}$ is defined to be copositive if the quadratic form $x^T\bm{\Lambda} x \geq 0$ for all positive vectors (x) in non-negative orthant\footnote{An orthant can be defined as an $n$-dimensional (Euclidean space) generalization of the quadrant and octant in two and three-dimensional space, respectively.} of} $\mathbb{R}_n$ (real coordinate vector space), \textit{i.e.} $\mathbb{R}^+_n$. The positive-definite matrices are a subset of the copositive matrices. A non-negative matrix $\Lambda$ of any order with its components $\lambda_{ij} \geq 0$ can be easily seen as a copositive matrix.
A symmetric matrix of order 2 is copositive iff \cite{HADELER198379,NADLER1992195}
\begin{equation} \label{eq.24}
  \l_{ii} \geq0, \quad \l_{jj} \geq0, \quad \text{and} \quad \l_{ij} + \sqrt{\l_{ii} \l_{jj}} \geq 0.
\end{equation}
where $i\neq j$. For the case of a three-order symmetric matrix, the copositive conditions read \cite{HADELER198379,CHANg_1994113}
\begin{eqnarray} \label{eq.25}
  \l_{ii} \geq 0, \quad \l_{jj} \geq0, \quad \l_{kk}\geq0, \nonumber\\
  \bar{\l}_{ij} \equiv \l_{ij} + \sqrt{\l_{ii} \l_{jj}} \geq 0, \nonumber\\
  \bar{\l}_{ik} \equiv \l_{ik} + \sqrt{\l_{ii} \l_{kk}} \geq 0, \nonumber\\
  \bar{\l}_{jk} \equiv \l_{jk} + \sqrt{\l_{jj} \l_{kk}} \geq 0, \nonumber\\
  \sum_{i,j,k} \l_{ij} \sqrt{\l_{kk}} +
  \sqrt{\l_{ii} \l_{jj} \l_{kk}} + \sqrt{2 \bar{\l}_{ij} \bar{\l}_{ik}
  \bar{\l}_{jk}} \geq 0.
\end{eqnarray}
where $i\neq j \neq k$. When dealing with a $4\times4$ symmetric matrix, we adopt the copositivity conditions derived in \cite{PING1993109}.
For the potential in Eq.\eqref{eq.1}, its quartic part can be written as
\begin{equation} \label{eq.18}
    \begin{split}
        V^{(4)} &= \l_1 (\eta^\d \eta)^2 + \l_2 (\eta^\d \eta)(H^\d H) + [\l_3 (\eta^\d H)^2 + \mathrm{H.c.}] + \l_4 (H^\d H)^2  \\
        & \quad + \frac{1}{4} \l_5 \Phi^4 + \l_6 (\Phi'^\d \Phi')^2 + \frac{\l_7}{2} H^\d H \Phi^2 + \l_{8} H^\d H (\Phi'^\d \Phi') + \frac{\l_9}{2} \Phi^2(\Phi'^\d \Phi') \\
        & \quad + \l_{10} (\eta^\d H)(H^\d \eta) + \l_{11} (\eta^\d \eta)(\Phi'^\d \Phi') + \l_{12} (\eta^\d \eta)\Phi^2.
    \end{split}
\end{equation}
In $(\eta,H,\Phi',\Phi)$ basis, their quartic coupling matrix take the following form,
\begin{equation} \label{eq.cm}
    \Lambda = \begin{pmatrix}
        \l_1 & \l_2 + \l_3 + \l_{10} & \l_{11} & \l_{12} \\
        \l_2 + \l_3 + \l_{10} & \l_4 & \l_8 & \frac{1}{2} \l_7 \\
        \l_{11} & \l_8 & \l_6 & \frac{1}{2} \l_9 \\
        \l_{12} & \frac{1}{2} \l_7 & \frac{1}{2} \l_9 & \l_5
    \end{pmatrix}.
\end{equation}
The very obvious copositivity conditions for the potential $V^{(4)}$ to be bounded from below are
\begin{equation}
    \l_1 \geq 0, \quad \l_4 \geq 0, \quad \l_5 \geq 0, \quad \l_6 \geq 0.
\end{equation}
On top of that, there are eight different cases related to the copositivity conditions depending on the signs of the off-diagonal elements. These will give rise to strong constraints on the parameters of the potential. We provide the complete sets of these constraints in Appendix \ref{app.A}.

\subsection{Good vacuum and global minimum} \label{sec.3.2}
Now, from the previous subsection, we have ensured that the scalar potential can be bounded from below with certain conditions on the quartic couplings. However, it is also essential to identify the 'trueness' of the electroweak vacuum for the model.
Various numerical methods have been developed to find the minima of functions or systems with multiple variables.
Below, we adopt the numerical procedure explained in Ref.\cite{BhupalDev:2018xya} and perform numerical minimization of the potential in Eq.\eqref{eq.1} with a one-order numerical method, namely, the Nelder-Mead algorithm. 
Before numerically minimizing the potential, we first propose the criteria of a good vacuum for symmetry breaking as follows,
\begin{align} \label{eq34}
    1. & \langle \eta \rangle = 0, \nonumber \\
    2. & \langle H \rangle \neq 0, \nonumber \\
    3. & \langle \Phi \rangle \neq 0, \nonumber \\
    4. & \langle \Phi^\prime \rangle \neq 0.
\end{align}
One can verify that if a minimum satisfies the above criteria, the desired VEVs alignment must take the form specified in Eq.\eqref{eq.2}.

For our numerical purposes, we begin by randomly assigning values to the potential parameters to observe the resulting VEV alignments. The quadratic, cubic, and quartic couplings are randomly generated from a uniform distribution within the interval \([-4\pi, 4\pi]\).\footnote{In this respect, quadratic and cubic couplings should be multiplied by the value of VEV to obtain the correct dimension.} However, since the randomly generated quartic couplings do not ensure the BfB condition of the potential, we implement a numerical procedure to verify this condition.

In our BfB check, we set the quadratic and cubic couplings to zero, leaving only the quartic terms. We then proceed with the minimization process using a numerical algorithm. During this minimization, there may be cases where \(V < 0\) is encountered, in which the corresponding samples are eliminated. Conversely, samples that pass this check undergo further minimization with non-zero quadratic and cubic couplings. Once we obtain a minimum, we check whether the result violates any of the four specified conditions in Eq.\eqref{eq34}. If a sample violates any of these criteria, it is categorized as type 1, 2, 3, or 4 based on the specific condition that was violated. On the contrary, if it meets all four conditions, it is labeled as type 5. Further details regarding the definitions of each type can be found in Table \ref{table:1}.
\begin{table}[!ht]
    \centering
    \begin{tabular}{|cp{0.3\linewidth}cccc|}
        \hline
         & & & & & \\[-0.5em]
        type & definition & $\langle\eta\rangle$ & $\langle H\rangle$ & $\langle\Phi\rangle$ & $\langle\Phi^\prime\rangle$ \\
         & & & & & \\[-0.5em]
        \hline
         & & & & & \\[-0.5em]
        1 & violates the first condition of Eq.\eqref{eq34}; $\langle \eta\rangle \neq 0$ & $\begin{pmatrix}
            0 \\ a_1
        \end{pmatrix}$ & $\begin{pmatrix}
            0 \\ a_2
        \end{pmatrix}$ & $a_3$ & $a_4$ \\
        & & $\begin{pmatrix}
            0 \\ a_1
        \end{pmatrix}$ & $\begin{pmatrix}
            0 \\ 0
        \end{pmatrix}$ & $a_3$ & $a_4$ \\
        & & & & & \\[-0.5em]
        & & $\begin{pmatrix}
            0 \\ a_1
        \end{pmatrix}$ & $\begin{pmatrix}
            0 \\ a_2
        \end{pmatrix}$ & 0 & $a_4$ \\
        & & & & & \\[-0.5em]
        & & $\dotsm$ & $\dotsm$ & $\dotsm$ & $\dotsm$ \\
        & & & & & \\[-0.5em]
        \hline
         & & & & & \\[-0.5em]
        2 & violates the second condition of Eq.\eqref{eq34}; $\langle H\rangle = 0$ & $\begin{pmatrix}
            0 \\ 0
        \end{pmatrix}$ & $\begin{pmatrix}
            0 \\ 0
        \end{pmatrix}$ & $a_3$ & $a_4$ \\
        & & $\begin{pmatrix}
            0 \\ 0
        \end{pmatrix}$ & $\begin{pmatrix}
            0 \\ 0
        \end{pmatrix}$ & 0 & $a_4$ \\
        & & & & & \\[-0.5em]
        & & $\begin{pmatrix}
            0 \\ 0
        \end{pmatrix}$ & $\begin{pmatrix}
            0 \\ 0
        \end{pmatrix}$ & $a_3$ & 0 \\
        & & & & & \\[-0.5em]
        & & $\begin{pmatrix}
            0 \\ 0
        \end{pmatrix}$ & $\begin{pmatrix}
            0 \\ 0
        \end{pmatrix}$ & 0 & 0 \\
        & & & & & \\[-0.5em]
        \hline
        & & & & & \\[-0.5em]
        3 & violates the third condition of Eq.\eqref{eq34}; $\langle \Phi\rangle = 0$ & $\begin{pmatrix}
            0 \\ 0
        \end{pmatrix}$ & $\begin{pmatrix}
            0 \\ a_2
        \end{pmatrix}$ & 0 & $a_4$ \\
        & & $\begin{pmatrix}
            0 \\ 0
        \end{pmatrix}$ & $\begin{pmatrix}
            0 \\ a_2
        \end{pmatrix}$ & 0 & 0 \\
        & & & & & \\[-0.5em]
        \hline
        & & & & & \\[-0.5em]
        4 & violates the fourth condition of Eq.\eqref{eq34}; $\langle \Phi^\prime \rangle = 0$ & $\begin{pmatrix}
            0 \\ 0
        \end{pmatrix}$ &$\begin{pmatrix}
            0 \\ a_2
        \end{pmatrix}$ & $a_3$ & 0 \\
        & & & & & \\
        \hline
        & & & & & \\[-0.5em]
        5 & satisfies Eq.\eqref{eq34} & $\begin{pmatrix}
            0 \\ 0
        \end{pmatrix}$ & $\begin{pmatrix}
            0 \\ a_2
        \end{pmatrix}$ & $a_3$ & $a_4$ \\
        & & & & & \\
        \hline
    \end{tabular}
    \caption{Five types of minima and the corresponding VEV alignments found in the numerical analysis. Each type is defined by a relation that violates the proposed good vacuum conditions in Eq.\eqref{eq34}, except for type 5. The VEV alignments are listed in the 3rd to 6th column where $a_i$ $(i=1,2,3,4)$ indicate nonzero values of the three fields.}
    \label{table:1}
\end{table}

\begin{figure}[t]
    \centering
    \includegraphics[width=0.5\linewidth]{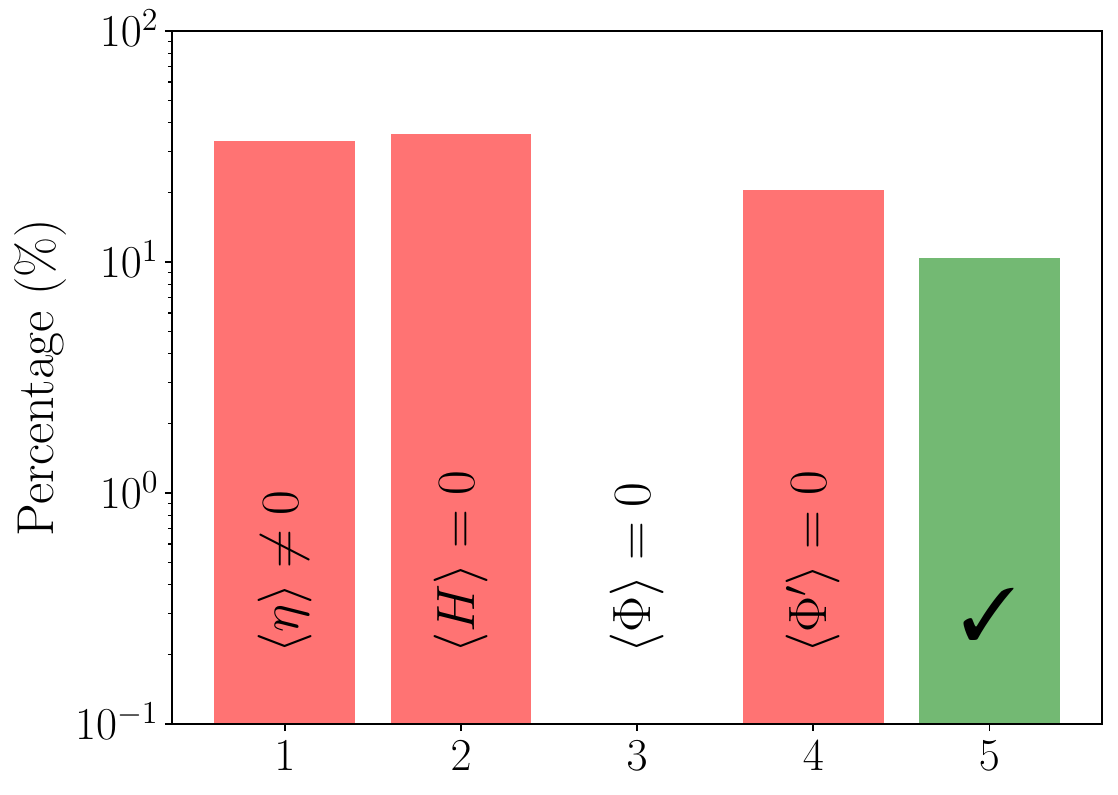}
    \caption{Percentages of the five types of vacuum found in the randomly generated samples. Type 1 to 5 cover around 33.4\%, 35.7\%, 0\%, 20.5\%, and 10.4\%, respectively. The definition of each type of the obtained vacuum is given in Table \ref{table:1}.}
    \label{fig:3}
\end{figure}

We apply the aforementioned procedure on a sample size of 2 million and present the minimization results for 47336 samples (all of which passed the BfB check) in Figure  \ref{fig:3}. Among the checked samples, 15803, 16906, 0, and 9696 are classified into types 1, 2, 3, and 4, respectively. In particular, 4931 samples fall into category type 5, where they meet the symmetry-breaking requirements outlined in Eq.~\eqref{eq34}. The type 5 category represents approximately 10.4\% of the total sample. This small percentage indicates that while the BfB conditions are necessary, they are not sufficient to achieve an optimally good vacuum.

\begin{figure}[t]
    \centering
    \begin{tabular}{cc}
    \includegraphics[width=8cm]{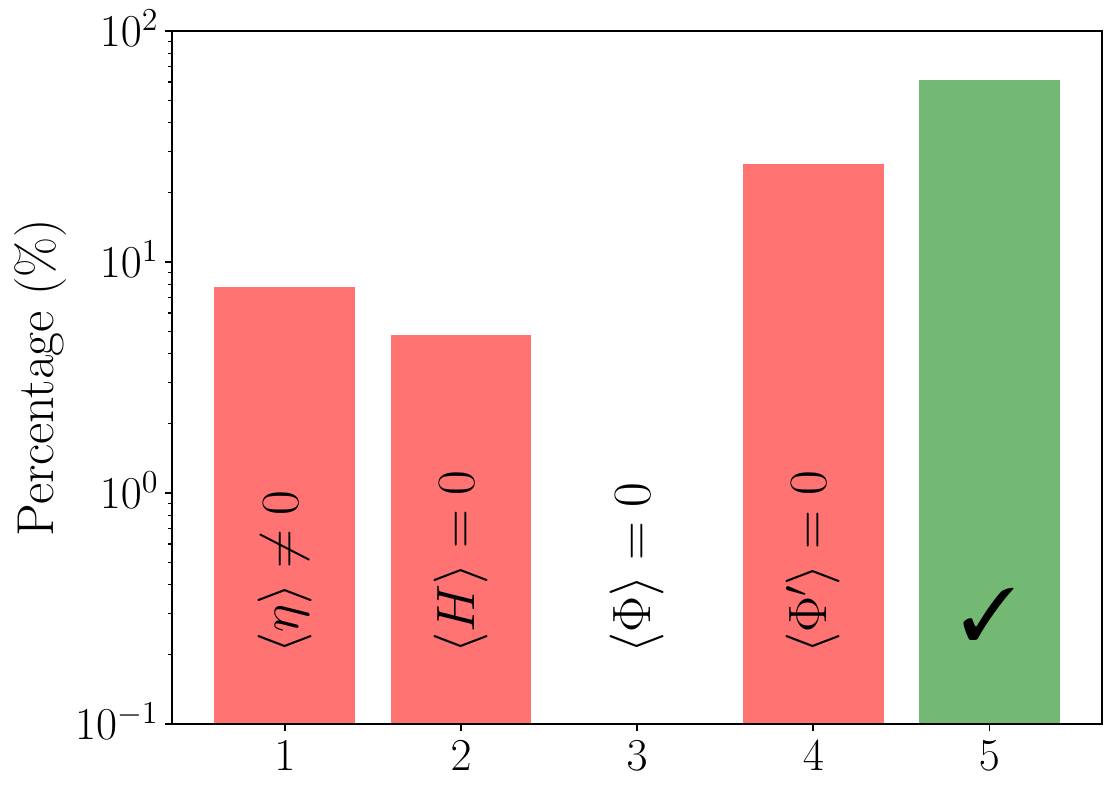} & \includegraphics[width=8cm]{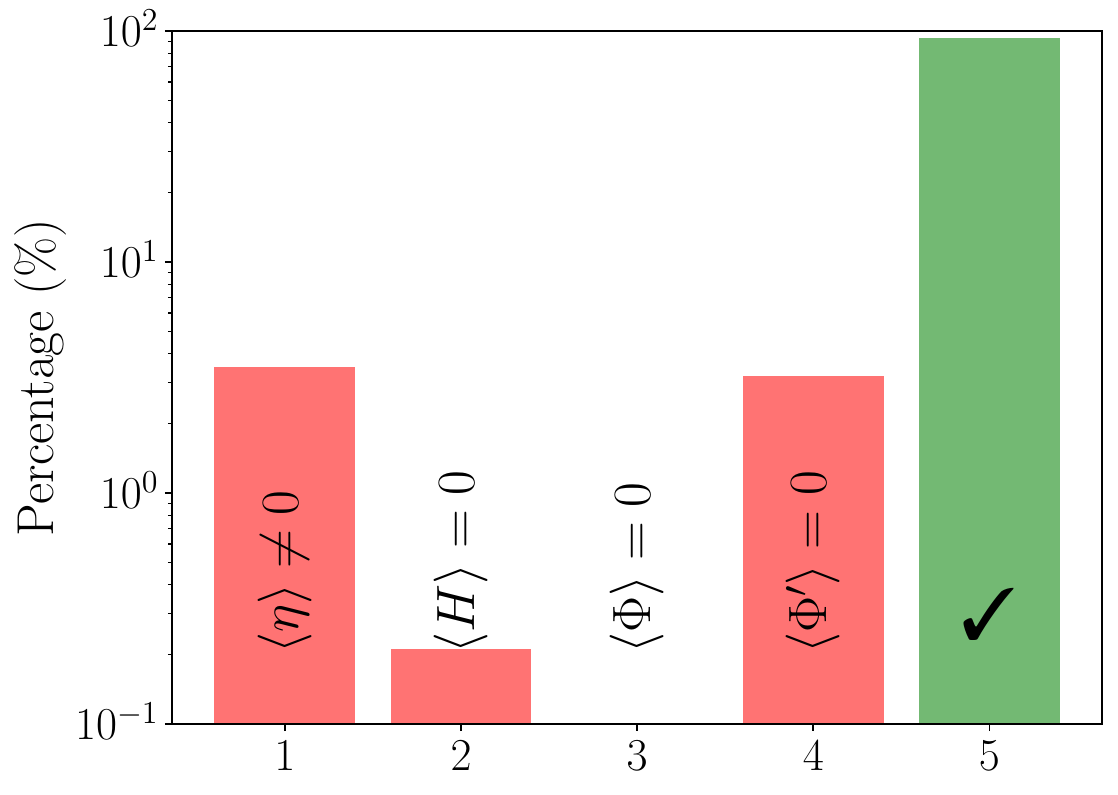}
    \end{tabular}
    \caption{Percentages of the five types of vacuum when the potential parameters are generated with additional constraints given in Eq.\eqref{eq41} (left panel) and Eq.\eqref{eq42} (right panel). In the left panel, the percentages are 7.8\%, 4.8\%, 0\%, 26.4\%, and 61.0\%. In the right panel, the percentages are 3.5\%, 0.2\%, 0\%, 3.2\%, and 93.1\%. All for types 1 to 5, respectively.} 
    \label{fig:4}
\end{figure}

To enhance the probability of obtaining a sample with a good vacuum (type 5), we incorporate the BfB conditions obtained in Sec.\ref{sec3} and choose the corresponding constraints as follows,
\begin{align} \label{eq41}
   & 0 \leq \mu_{1,2,3,5}^2 \leq 4\pi v_2^2,  \nonumber\\
   & 0 \leq \l_{1,4,5,6} \leq 4\pi.
\end{align}
By enforcing the above constraints in the generation of samples, we repeat the numerical minimization (this time with 100k data) and obtain the probability enhancement of type 5 up to $61.0\%$, which is shown by the left panel of Figure~\ref{fig:4}. This percentage can be further enhanced by imposing more constraints. In addition to the above constraints, we put in the following new constraints
\begin{eqnarray} \label{eq42}
     -0.05\pi v_2 \leq \mu_{4,6,7} \leq 0.05\pi v_2, \qquad
     -0.2\pi \leq \l_{7,8,9} \leq 0.2\pi, \nonumber \\
     -0.1\pi \leq \l_{2,11,12} \leq 0.1\pi, \qquad
     -0.1\pi \leq \l_{3,10} \leq 0,
\end{eqnarray}
and obtain a higher percentage for type 5, which is around $93.1\%$. Note that, in the above expression, $\l_{3,10} $ are chosen within those values to keep the physical charged Higgs mass positive. The result is presented in the right panel of Figure~\ref{fig:4}. As shown in the plot, the increasing percentage of type 5 implies that most of the samples, being generated with both constraints in Eqs.\eqref{eq41} and \eqref{eq42}, have good vacua. In any case, they give rise to the occurrence of spontaneous symmetry breaking with correct VEV alignment (see Eq.\eqref{eq34}). Nevertheless, it is worth noting that the aforementioned conditions provide a necessary but not sufficient condition for obtaining an optimally good vacuum, as indicated by the statistical results.

\begin{figure}[h!]
    \centering
    \begin{tabular}{cc}
        \includegraphics[width=6.5cm]{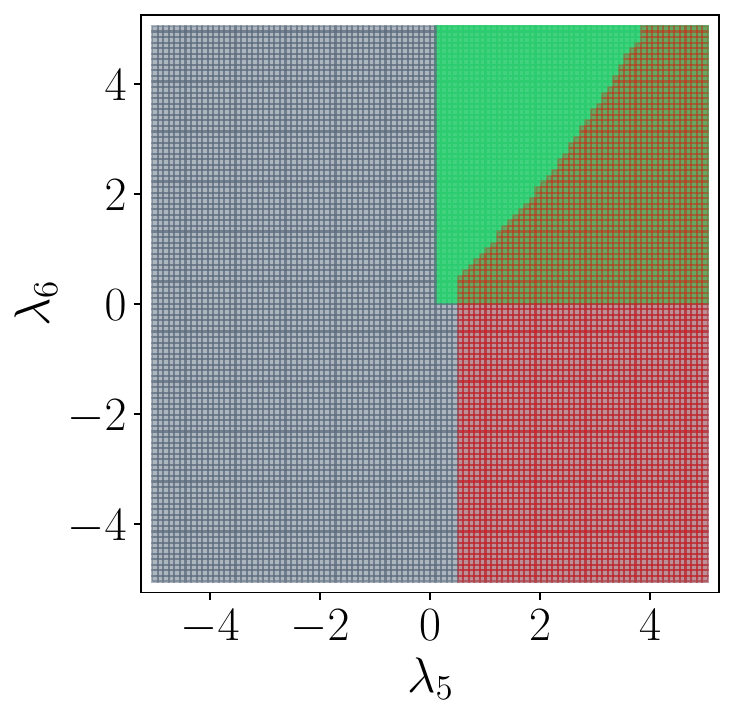} & \includegraphics[width=6.5cm]{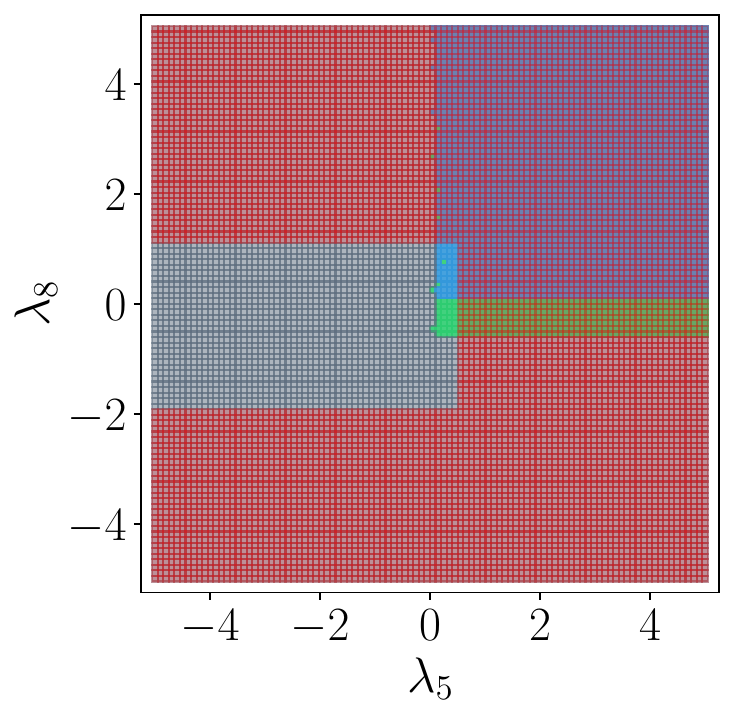} 
        \\
        \multicolumn{2}{c}{\includegraphics[width=6.5cm]{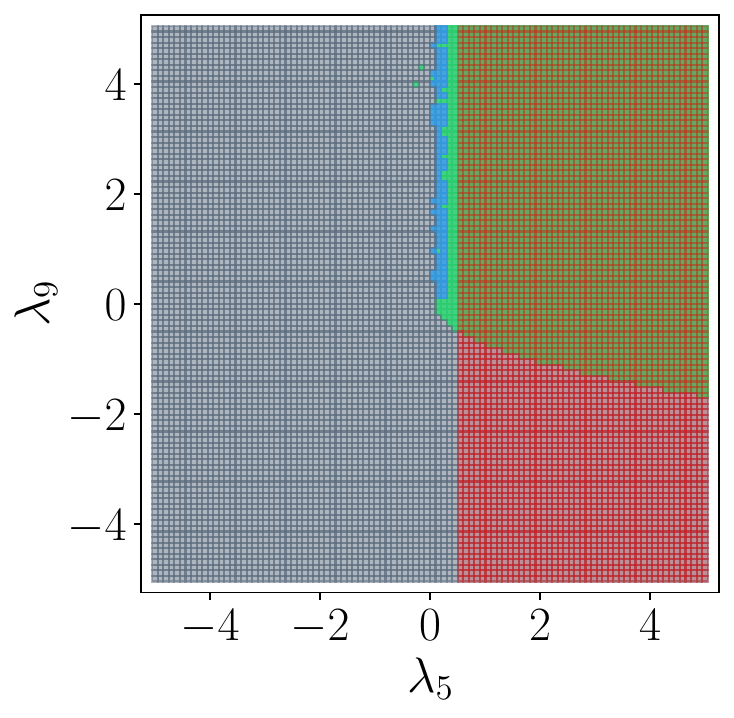}}
    \end{tabular}
    \caption{The numerical minimization of the scalar potential in Eq.\eqref{eq.1} is plotted for different pairs of quartic couplings $\l_i\,(i=5,6,8,9)$. Other parameters, besides the ones shown as axes, are fixed at the benchmark value in Eq.\eqref{eq43}.
    The gray region violates the BfB condition. The blue region satisfies the BfB condition but does not include a type 5 vacuum. The green region represents a BfB potential and fulfills the type 5 vacuum. The red region indicates all data points exceeding the perturbative limit $4\pi$.}
    \label{fig:5}
\end{figure}

Based on the above statistical results, we selected the following benchmark point to investigate the potential parameters of the model and their alignment with the observed conditions. They are given by,
\begin{equation} \label{eq43}
\begin{split}
    \mu_1^2, \mu^2_2, \mu^2_3, \mu^2_5 &\equiv (0.2,0.1, 0.02, 0.2) v^2_2, \\
    \mu_4, \mu_6, \mu_7 &\equiv (0.01, -10^{-4}, -10^{-4}) v_2,  \\
    \l_1,\l_2, \l_3, \l_4 &\equiv (10^{-9},10^{-11}, -10^{-8}, 0.129),  \\
    \l_5,\l_6,\l_7, \l_8 &\equiv (0.01,0.15, -0.01, -0.1),  \\
    \l_9, \l_{10},\l_{11},\l_{12} &\equiv (-10^{-3},-10^{-7}, 10^{-10}, 10^{-10}).
    \end{split}
\end{equation}
The above benchmark points satisfy the conditions in Eq.\eqref{eq41} and Eq.\eqref{eq42}, which allow us to present three plots of constraints on a pair of quartic couplings $\l_i$ ($i=5,6,8,9$) while the other parameters are fixed. These constraints are shown in Figure~\ref{fig:5}.

Figure~\ref{fig:5} presents the constraints of $\l_i$ ($i=6,8,9$) as a function of $\l_5$ in the form of data distribution of the minimized potential. In this figure, each plot is generated on a $100\times100$ grid within the range of (-5, 5), where each pixel has a size of $0.1\times0.1$.
The green region illustrates the existence of data with a bounded minimum owing to the condition of type 5, facilitating the correct symmetry breaking. The blue region indicates more general conditions where the potential is bounded from below but does not result in the desired symmetry breaking. The grey region represents unbounded minima that violate the BfB conditions. It is important to note that the selected points do not fully meet the criteria necessary for successful symmetry breaking, and the positive value of $\l_5$ indeed leads to the type 5 vacuum, aligning with the findings discussed in the previous section.

\section{Renormalization group equations and perturbative constraints} \label{sec4}
In this section, we analyze the evolution of the quartic couplings of the scalar potential using the renormalization group equations (RGEs). We aim to study how these couplings behave as the energy scale increases, starting from specific initial conditions. While the initial values of the couplings may satisfy the previous constraints—the bounded-from-below condition and the presence of a global minimum—it is still possible for the couplings to grow rapidly and exceed the perturbative limit $4\pi$. 

To study this behavior systematically, we employ numerical simulations with a similar setting as in Section \ref{sec.3.2}. We first derive the RGEs of the quartic couplings using the PyR@TE package \cite{Sartore:2020gou} and the RGBeta package\footnote{We have also used the RGBeta package for the purpose of checking consistency and comparison only.} \cite{Thomsen:2021ncy} within Mathematica.
These tools allow us to derive the beta functions for all relevant couplings, including:
\begin{itemize}
    \item Scalar quartic couplings; $\lambda_i$ ($i=1,2,...,12$),
    \item Gauge couplings; $g_1$, $g_2$, and $g_3$,
    \item Yukawa couplings\footnote{In Ref.\cite{Dutta:2022knf}, the Yukawa interactions such as $\lambda_{D}\bar{\chi} \chi \Phi$, $y_{\nu_{R}} \bar{L}\tilde{X}\nu_{R}$, and $\lambda_{\chi}\bar{\chi} L \eta$ are introduced where $L$ stands for the SM lepton fields. The terms with couplings, $\lambda_{\chi}$ and $y_{\nu_{R}}$ give rise to the dimension eight and five operators, respectively. For the detailed discussion regarding this, one can refer to Ref.\cite{Dutta:2022knf}.}; $\lambda_D$, $\lambda_\chi$, $y_{\nu_R}$, and $y_t$ (the SM top quark Yukawa coupling).
\end{itemize}
The scalar quartic beta functions up to one-loop order are given by
\begin{align}
    16\pi^2 \frac{d\lambda_1}{d\mu} &= 24 \lambda_1^{2} + 2 \lambda_2^{2} + 4 \lambda_3^{2} + 2 \lambda_{10} \lambda_2 + \lambda_{10}^{2} + \lambda_{11}^{2} + 2 \lambda_{12}^{2} - \left(3 g_1^{2} + 9 g_2^{2}\right) \lambda_1 + \frac{3}{8} g_1^{4} + \frac{3}{4} g_1^{2} g_2^{2} \nonumber \\
    & \quad + \frac{9}{8} g_2^{4} + 4 \lambda_1 \lambda_\chi^2 - 2 \lambda_\chi^4, \label{eqbetal1} \\
    16\pi^2 \frac{d\lambda_2}{d\mu} &= 4 \lambda_2^{2} + 12 \lambda_1 \lambda_2 + 4 \lambda_1 \lambda_{10} +  12 \lambda_2 \lambda_4 + 8 \lambda_3^{2} + 4 \lambda_{10} \lambda_4 + 2 \lambda_{10}^{2} + 2 \lambda_{11} \lambda_8 + 2 \lambda_{12} \lambda_7 \nonumber \\
    & \quad - \left(3 g_1^{2} + 9 g_2^{2}\right) \lambda_2 + \frac{3}{4} g_1^{4} - \frac{3}{2} g_1^{2} g_2^{2} + \frac{9}{4} g_2^{4} + 6 \lambda_2 y_t^2 + 2 \lambda_2 \lambda_\chi^2, \\
    16\pi^2 \frac{d\lambda_3}{d\mu} &= 4 \lambda_1 \lambda_3 + 8 \lambda_2 \lambda_3 + 4 \lambda_3 \lambda_4 + 12 \lambda_{10} \lambda_3 - \left(3 g_1^{2} + 9 g_2^{2}\right) \lambda_3 + 6 \lambda_3 y_t^2 + 2 \lambda_3 \lambda_\chi^2, \\
    16\pi^2 \frac{d\lambda_4}{d\mu} &= 24 \lambda_4^2 + 2\lambda_2^2 + 4\lambda_3^2 + \frac{1}{2}\lambda_7^2 + \lambda_8^2 + 2 \lambda_{10} \lambda_2 + \lambda_{10}^{2} - \left( 3 g_1^{2} + 9 g_2^{2}\right) \lambda_4 + \frac{3}{8} g_1^{4} \nonumber \\
    & \quad + \frac{3}{4} g_1^{2} g_2^{2} + \frac{9}{8} g_2^{4} + 12 \lambda_4 y_t^2 - 6 y_t^4, \\
    16\pi^2 \frac{d\lambda_5}{d\mu} &= 18 \lambda_5^{2} + 2 \lambda_7^{2} + \lambda_9^{2} + 8 \lambda_{12}^{2} + 16 \lambda_5 \lambda_D^{2} - 64 \lambda_D^{4}, \\
    16\pi^2 \frac{d\lambda_6}{d\mu} &= 20 \lambda_6^{2} + 2 \lambda_8^{2} + \frac{1}{2} \lambda_9^{2} + 2 \lambda_{11}^{2}, \\
    16\pi^2 \frac{d\lambda_7}{d\mu} &= 4 \lambda_7^{2} + 12 \lambda_4 \lambda_7 + 6 \lambda_5 \lambda_7 + 2 \lambda_8 \lambda_9 + 4 \lambda_{10} \lambda_{12} + 8 \lambda_{12} \lambda_2 - \left(\frac{3}{2} g_1^{2} + \frac{9}{2} g_2^{2}\right) \lambda_7 \nonumber \\
    & \quad + 8 \lambda_7 \lambda_D^{2} + 6 \lambda_7 y_t^2, \\
    16\pi^2 \frac{d\lambda_8}{d\mu} &= 4 \lambda_8^{2} + 12 \lambda_4 \lambda_8 + 8 \lambda_6 \lambda_8 + \lambda_7 \lambda_9 + 2 \lambda_{10} \lambda_{11} + 4 \lambda_{11} \lambda_2 -  \left(\frac{3}{2} g_1^{2} +  \frac{9}{2} g_2^{2}\right) \lambda_8 \nonumber \\
    & \quad + 6 \lambda_8 y_t^2, \\
    16\pi^2 \frac{d\lambda_9}{d\mu} &= 4 \lambda_9^{2} + 6 \lambda_5 \lambda_9 + 8 \lambda_6 \lambda_9 + 4 \lambda_7 \lambda_8 + 8 \lambda_{11} \lambda_{12} + 8 \lambda_9 \lambda_D^{2}, \\
    16\pi^2 \frac{d\lambda_{10}}{d\mu} &=  4 \lambda_{10}^{2} + 4 \lambda_1 \lambda_{10} + 8 \lambda_2 \lambda_{10} + 32 \lambda_3^{2} + 4 \lambda_4 \lambda_{10} - \left(3 g_1^{2} + 9 g_2^{2}\right)\lambda_{10} + 3 g_1^{2} g_2^{2} \nonumber \\
    &\quad + 6 \lambda_{10} y_t^2 + 2 \lambda_{10} \lambda_\chi^2, \\
    16\pi^2 \frac{d\lambda_{11}}{d\mu} &= 4 \lambda_{11}^{2} + 12 \lambda_1 \lambda_{11} + 4 \lambda_2 \lambda_8 + 2 \lambda_{10} \lambda_8 + 8 \lambda_{11} \lambda_6 + 2 \lambda_{12} \lambda_9 -  \left(\frac{3}{2} g_1^{2} + \frac{9}{2} g_2^{2}\right) \lambda_{11} \nonumber\\
    & \quad + 2 \lambda_{11} \lambda_\chi^2, \\
    16\pi^2 \frac{d\lambda_{12}}{d\mu} &= 8 \lambda_{12}^{2} + 12 \lambda_1 \lambda_{12} + 2 \lambda_2 \lambda_7 + \lambda_{10} \lambda_7 + \lambda_{11} \lambda_9 + 6 \lambda_{12} \lambda_5 - \left(\frac{3}{2} g_1^{2} + \frac{9}{2} g_2^{2}\right) \lambda_{12} \nonumber \\
    & \quad + 8 \lambda_{12} \lambda_D^{2} + 2 \lambda_{12} \lambda_\chi^2 - 16 \lambda_D^{2} \lambda_\chi^2, \label{eqbetal12}
\end{align}
in which we have omitted the contributions of the SM bottom quark $y_b$ and tau lepton $y_\tau$ Yukawa couplings because of their negligibly small values. The beta function equations at the one-loop level for the gauge and the Yukawa couplings are provided in Appendix \ref{app.B}.

We then evolve the RGEs numerically from the electroweak scale, characterized by the top quark mass ($M_t$) up to the Planck scale ($M_{\text{Pl}}=1.22\times10^{19}~\text{GeV}$), to further investigate the behavior of the couplings over this range of energy scales.
This logarithmic range spans many orders of magnitude and allows us to probe the theoretical consistency of the model at both low and high energies. 

To explore the parameter space, we vary the initial values of the quartic couplings, focusing particularly on pairs involving $\lambda_5$ and $\lambda_i$ $(i=6,8,9)$. These pairs are chosen because of their significant influence on the running behavior of the scalar potential. The initial values are sampled in a grid-like approach, where $\lambda_5$ and $\lambda_i$ are varied from -5 to 5 in increments of 0.1, consistent with the parameters set in the previous section. The other quartic couplings, except for those in the chosen pairs, are fixed according to predefined benchmark values in Eq.~\eqref{eq43}. We use the initial values of the gauge couplings and the Yukawa couplings at energy scale $\mu = M_t$, as provided in Table \ref{table:2} \cite{DuttaBanik:2020jrj}, where $\lambda_H\equiv\lambda_4$.
\begin{table}[t]
    \centering
    \begin{tabular}{cccccc}
    \hline \hline
    & & & & & \\[-0.9em]
    Scale & $\lambda_H$ & $y_t$ & $g_1$ & $g_2$ & $g_3$ \\
    & & & & & \\[-0.9em]
    \hline
    & & & & & \\[-0.9em]
    $\mu = M_t$ & 0.125932 & 0.936100 & 0.357606 & 0.648216 & 1.166550\\
    & & & & & \\[-0.9em]
    \hline \hline
    \end{tabular}
    \caption{The values of the relevant SM couplings at energy scale $\mu=M_t = 173.2~\text{GeV}$ with $M_h = 125.09~\text{GeV}$ and $\alpha_S (M_Z) = 0.1184~\text{GeV}$ \cite{DuttaBanik:2020jrj}.}
    \label{table:2}
\end{table}
These values are aligned with Ref.\cite{Braathen:2017jvs} and evaluated by using the boundary values given in \cite{Buttazzo:2013uya}. We also specifically choose $\lambda_D \geq 0.03,~\l_\chi = 0.01,~\text{and}~ y_{\nu_R} = 0.01$. Here, the value of $\lambda_D$ is consistent with the lower limit value in the analysis for the dark fermion $\chi$ self-interaction that has been done in \cite{Dutta:2022knf}.

\begin{figure}
    \centering
    \includegraphics[width=0.6\linewidth]{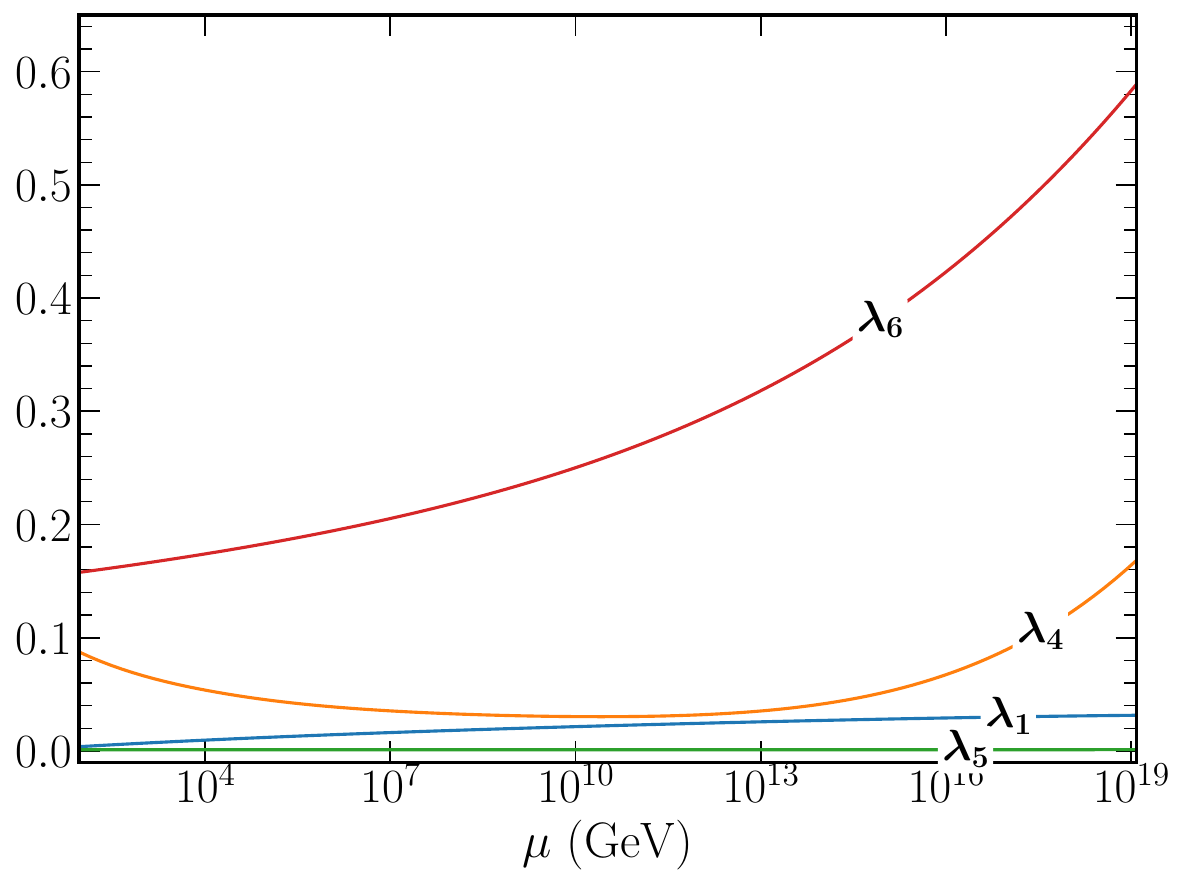}
    \caption{RG running of the chosen quartic couplings $\lambda_i\,(i=1,4,5,6)$ of the scalar potential (Eq.\eqref{eq.1}) with initial values given in Table~\ref{table:2} and BP1 in Table \ref{table:3}. The other couplings, besides $\lambda_i$, overlap within the region below $0.05$ and are not shown for clarity.}
    \label{fig:rge}
\end{figure}

Figure~\ref{fig:rge} shows the RG evolution of the self-interaction quartic couplings $\lambda\,(i=1,4,5,6)$. The initial values of quartic couplings satisfy all previous constraints, given by Eq.\eqref{eq43} with $\lambda_5 =10^{-3}$ and $\lambda_8 = -10^{-2}$ (further referred to as BP1 in Table \ref{table:3}). The other couplings $\lambda_j\,(j=2,3,7,8,9,10,11,12)$ exhibit overlapping behavior within the ($\leq 0.05$) region and therefore are not shown for clarity. From the figure, we note that the SM-like Higgs quartic coupling remains positive, $\lambda_4 > 0$, throughout the energy range. The evolution demonstrates that the couplings remain perturbative up to the Planck scale, ensuring a stable and well-behaved scalar potential.

\section{Parameter space} \label{sec5}
In this section, utilizing the results from the previous sections, we perform a numerical study for the parameter space of the model within the allowed ranges of the model's couplings. For this purpose, we take into account theoretical and experimental constraints, including the bounded-from-below conditions, global minimum constraint, the perturbative constraint of the quartic couplings up to the Planck scale, and the Higgs invisible decay. Throughout this numerical study, we will set $v_2 \simeq 246$ GeV, and $m_{H_2} \simeq 125$ GeV for the SM Higgs-like field. 

\begin{figure}
    \centering
    \includegraphics[width=0.5\linewidth]{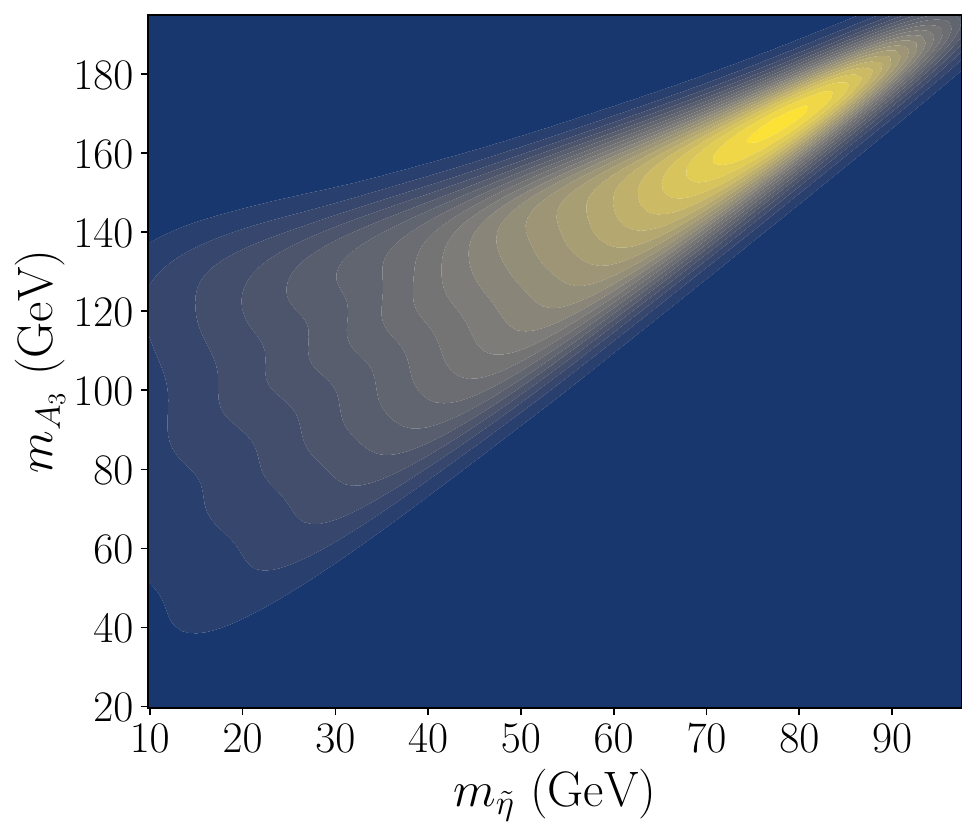}
    \caption{Distribution of parameter points in the $m_{\tilde{\eta}}-m_{A_3}$ plane satisfying all theoretical constraints. The color shading indicates the density of viable solutions, with yellow corresponding to higher concentrations of allowed points and blue to lower concentrations. The overlaid contours highlight levels of constant density, while the darkest region outside the contours is excluded.}
    \label{fig:mhc-mA3}
\end{figure}

We first explore the mass spectra for the charged scalar $\tilde{\eta}$ and the massive pseudoscalar $A_3$.
Figure \ref{fig:mhc-mA3} shows the distribution of viable parameters within the $(m_{\tilde{\eta}},m_{A_3})$ plane. We use the benchmark values given in Eq.\eqref{eq43}, except that the values of {$\lambda_{3} \in (-0.1\pi,0)$ and $\lambda_{10} \in (-0.1\pi,0.1\pi)$} are varied, then map the solutions into physical masses. The color shading represents the density of parameter points that satisfy all theoretical constraints. Brighter regions (yellow) correspond to a higher concentration of allowed solutions, whereas darker regions (blue) indicate a lower density of viable points. The overlaid contours mark lines of constant density, and no viable points are found outside the contours (darkest blue), implying that these regions are excluded. This density plot structure further highlights that the majority of viable points cluster in the higher-mass region of the band, whereas the lower-mass region is more sparsely populated.

{Next, we select five benchmark points that satisfy the aforementioned theoretical constraints from Figure \ref{fig:mhc-mA3} and summarize them in Table \ref{table:3}. These BPs differ in their scalar couplings and mass spectra, offering different mixing scenarios and decay possibilities. 
By using these BPs, we confirm that all of the scalar masses are positive, and we show the results in Table \ref{table:4}.}

\begin{table}[t]
    \centering
    \resizebox{\textwidth}{!}{%
    \begin{tabular}{cccccccccccc}
    \hline \hline
    & & & & & & & & & & & \\[-0.9em]
    Benchmark & $\lambda_1$ & $\lambda_2$ & $\lambda_{3}$ & $\lambda_5$ & $\lambda_6$ & $\lambda_7$ & $\lambda_8$ & $\lambda_9$ & $\lambda_{10}$ & $\lambda_{11}$ & $\lambda_{12}$\\
    & & & & & & & & & & \\[-0.9em]
    \hline
    & & & & & & & & & & \\[-0.9em]
    BP1 & $10^{-9}$ & $10^{-11}$ & $-0.102$ & $10^{-3}$ & $0.15$ & $-0.01$ & $-0.01$ & $-10^{-3}$ &  $0.130$ & $10^{-10}$ & $10^{-10}$ \\
    & & & & & & & & & & &\\[-0.9em]
    BP2 & $10^{-11}$ & $5\times10^{-6}$ & $-0.187$  & $0.15$ & $0.01$ & $-0.2$ & $-1.5\times 10^{-3}$ & $-0.1$ & $0.238$ & $10^{-8}$ & $10^{-9}$ \\
    & & & & & & & & & & \\[-0.9em]
    BP3 & $6\times 10^{-8}$ & $10^{-9}$ & $-0.254$ & $0.02$ & $0.40$ & $-0.2$ & $-0.04$ & $-0.03$ & $0.314$ & $10^{-6}$ & $10^{-6}$\\
    & & & & & & & & & & \\[-0.9em]
    BP4 & $10^{-4}$ & $10^{-6}$ & $-0.218$ & $0.93$ & $0.08$ & $-10^{-3}$ & $-0.02$ & $-0.2$ & $0.225$ & $10^{-7}$ & $10^{-9}$ \\
    & & & & & & & & & & \\[-0.9em]
    BP5 & $10^{-8}$ & $10^{-9}$ & $-0.269$ & $0.01$ & $1.5\times 10^{-2}$ & $-10^{-3}$ & $-10^{-4}$ & $-10^{-3}$ & $0.269$ & $10^{-8}$ & $10^{-8}$ \\
    \hline \hline
    \end{tabular}%
    }
    \caption{The benchmark points satisfy all constraints with a fixed $\lambda_4\equiv \lambda_H = 0.125932$ \cite{DuttaBanik:2020jrj}.}
    \label{table:3}
\end{table}

\begin{table}[h!]
    \centering
    \resizebox{\textwidth}{!}{%
    \begin{tabular}{ccccccc}
        \hline \hline
         & & & & & & \\[-0.9em]
         Benchmark & $\bar{m}_{h_4}$ (GeV) & $m_{\tilde{\eta}}$ (GeV) & $m_{A_3}$ (GeV) & $m_{H_1}$ (GeV) & $m_{H_2}$ (GeV) & $m_{H_3}$ (GeV)  \\
         & & & & & & \\[-0.9em]
         \hline
         & & & & & & \\[-0.9em]
         BP1 & 99.473 & 46.994 & 110.862  & 547.742 & 124.868 & 30.775 \\
         & & & & & & \\[-0.9em]
         BP2 & 89.300 & 64.256 & 150.534 & 141.706 & 124.920 & 30.078 \\
         & & & & & & \\[-0.9em]
         BP3 & 79.035 & 76.532 & 175.288 & 894.496 & 124.527 & 30.556 \\
         & & & & & & \\[-0.9em]
         BP4 & 75.300 & 80.208 & 162.792 & 400.241 & 124.282 & 33.272 \\
         & & & & & & \\[-0.9em]
         BP5 & 62.782 & 90.341 & 180.683 & 173.205 & 124.952 & 30.805 \\
         \hline \hline
    \end{tabular}%
    }
    \caption{The scalar masses corresponding to five benchmarks.}
    \label{table:4}
\end{table}

Furthermore, we investigate the experimental constraints, particularly the upper bounds on the invisible decay branching ratio of the Higgs boson provided by ATLAS and CMS. In our model, the SM Higgs-like field $H_2$ can decay invisibly into any of a pair of dark fermions $\chi$, a pair of physically massless pseudoscalar fields $A_1$, a pair of lighter scalar fields $H_3$, or even a pair of massive pseudoscalar $A_3$. The decay width for the channel $H_2 \rightarrow \chi \bar{\chi}$ is given by
\begin{align}
    \Gamma_{H_2 \rightarrow \chi \bar{\chi}} = \lambda_D^2 (\mathcal{O}^T_{32})^2 \frac{m_{H_2}}{8 \pi} \left(1 - \frac{4 m^2_{\chi}}{m_{H_2}^2} \right)^{\frac{3}{2}},
\end{align}
where the mixing factor, $\mathcal{O}^T_{32} = \cos{\beta_2} \sin{\beta_3}$ (see Eq. \eqref{eq14}), corresponds to the contribution from the Yukawa interaction term between dark matter $\chi$ and scalar $h_3$: $\lambda_D \bar{\chi} \chi h_3$.

Another possible decay channel $H_2\rightarrow A_1 A_1$ is generated from the interaction term $\frac{1}{2} \l_8 v_2 h_2 \phi_1^2$  and its corresponding decay width is computed as follows,
\begin{equation}
    \Gamma_{H_2\rightarrow A_1 A_1} = (\lambda_8 v_2 \mathcal{O}_{22}^T)^2 \frac{m_{H_2}}{32 \pi}  \left( 1 - \frac{4 m^2_{A_1}}{m_{H_2}^2} \right)^{\frac{3}{2}},
\end{equation}
where $\mathcal{O}_{22}^T$ could be extracted from the mixing matrix in Eq.\eqref{eq14}. If kinematically allowed, the decay of $H_2$ into a pair of lighter scalars $H_3$ is possibly mediated through cubic and quartic interactions of the scalars. The corresponding decay width for the second channel $H_2 \rightarrow H_3 H_3$ is
 \begin{equation}
    \Gamma_{H_2 \rightarrow H_3 H_3} = \lambda_{h_2 h_3 h_3}^2 \frac{m_{H_2}}{8
    \pi} \left( 1 - \frac{4 m^2_{H_3}}{m_{H_2}^2}\right)^{\frac{3}{2}},
\end{equation}
where $\lambda_{h_2 h_3 h_3}$ represents the effective coupling constant involving various mixing factor terms and it is explicitly given by
\begin{align}
    \lambda_{h_2 h_3 h_3} &= 3 \lambda_4 v_2 \mathcal{O}_{22}^T (\mathcal{O}_{23}^T \mathcal{})^2 + 3 \lambda_5 v_3 \mathcal{O}_{32}^T {(\mathcal{O}_{33}^T)^2}  + 3 \lambda_6 v_1 \mathcal{O}_{12}^T (\mathcal{O}_{13}^T)^2 + \mu_4 \mathcal{O}_{32}^T (\mathcal{O}_{33}^T)^2 \nonumber \\
    & \quad + \lambda_7 \left( v_2 \mathcal{O}_{23}^T \mathcal{O}_{33}^T \mathcal{O}_{32}^T + \frac{v_2}{2} \mathcal{O}_{22}^T (\mathcal{O}_{33}^T)^2 + \frac{v_3}{2} (\mathcal{O}_{23}^T)^2 \mathcal{O}_{32}^T + v_3  \mathcal{O}_{23}^T \mathcal{O}_{33}^T \mathcal{O}_{22}^T \right) \nonumber\\
    & \quad + \lambda_8 \left( \frac{v_2}{2} \mathcal{O}_{22}^T (\mathcal{O}_{13}^T)^2 + v_2  \mathcal{O}_{12}^T \mathcal{O}_{13}^T \mathcal{O}_{23}^T + v_1 \mathcal{O}_{22}^T \mathcal{O}_{23}^T \mathcal{O}_{13}^T + \frac{v_1}{2} \mathcal{O}_{12}^T (\mathcal{O}_{23}^T)^2 \right) \nonumber \\
    & \quad + \lambda_9 \left( v_3  \mathcal{O}_{12}^T \mathcal{O}_{13}^T \mathcal{O}_{33}^T + \frac{v_3}{2} \mathcal{O}_{32}^T (\mathcal{O}_{13}^T)^2 + v_1 \mathcal{O}_{32}^T \mathcal{O}_{33}^T \mathcal{O}_{13}^T + \frac{v_1}{2} \mathcal{O}_{12}^T (\mathcal{O}_{33}^T)^2 \right)  \nonumber \\
    & \quad  + \frac{\mu_6}{2\sqrt{2}} ((\mathcal{O}_{23}^T)^2 \mathcal{O}_{32}^T + 2 \mathcal{O}_{23}^T \mathcal{O}_{33}^T \mathcal{O}_{22}^T) + \frac{\mu_7}{2 \sqrt{2}}  (2 \mathcal{O}_{12}^T \mathcal{O}_{13}^T \mathcal{O}_{33}^T + \mathcal{O}_{32}^T (\mathcal{O}_{13}^T)^2).
\end{align}

Further, a decay channel $H_2\rightarrow A_3 A_3$ may also open via an interaction term, $\lambda_{23} v_2 h_2 \phi_4^2$, with $\lambda_{23}\equiv \left(\frac{1}{2}\lambda_2+\lambda_3+\frac{1}{2}\lambda_{10}\right)$. The corresponding decay width is calculated as
\begin{align}
\Gamma_{H_2 \rightarrow A_3 A_3} = \left(\lambda_{23} v_2 \mathcal{O}_{22}^T\right)^2 \frac{m_{H_2}}{8\pi}\left(1-\frac{4m_{A_3}^2}{m_{H_2}^2}\right)^{\frac{3}{2}}.
\end{align}
{Note that, for the value of BP5 given in Table \ref{table:3}, this channel being kinematically not allowed}.

The total invisible branching ratio of the Higgs-like boson $H_2$ is then calculated as
\begin{equation}
    \text{BR}_{H_2 \rightarrow \text{inv}} = \frac{\Gamma_{H_2\rightarrow \text{inv}}}{4.4\,\text{MeV} + \Gamma_{H_2\rightarrow \text{inv}}},
\end{equation}
where $4.4 \ \text{MeV}$ \cite{ATLAS:2023dnm} is the current measurement of the SM Higgs boson's total width. The Higgs invisible branching ratio is constrained by the latest measurements by ATLAS \cite{ATLAS:2023tkt} ($BR\leq 0.107$) and by CMS \cite{CMS:2023sdw} ($BR\leq 0.15$).
\begin{figure}
    \centering
    \includegraphics[width=0.45\linewidth]{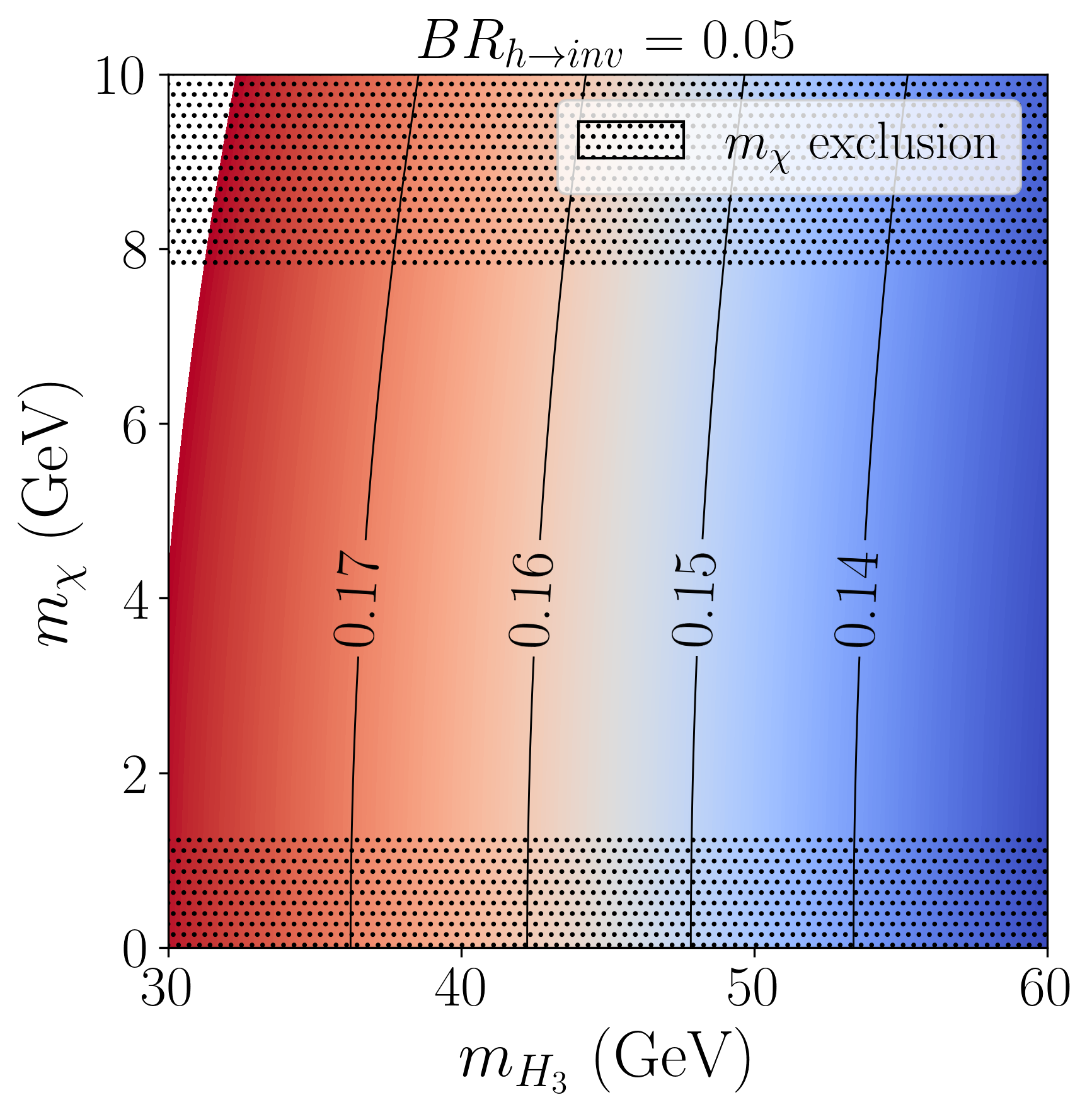}
    \caption{Higgs invisible decay constraint on the $m_{H_3}-m_\chi$ plane. The line contours show the value of $\lambda_D$, whereas the dotted region represents the excluded value of the dark fermion mass $m_\chi$: $m_\chi<1.3$ GeV and $m_\chi>7.76$ GeV \cite{Dutta:2022knf}.}
    \label{fig:BR-mass}
\end{figure}

To visualize allowed regions of the model, we perform numerical scans over physical parameters that control the invisible decay rates.
In Figure~\ref{fig:BR-mass}, we present the parameter space on the $(m_{H_3}, m_\chi)$ plane while taking $\text{BR}_{H_2 \rightarrow \text{inv}}$ to be $5\%$.
The contour lines represent values of the Yukawa coupling $\lambda_D$, which mediates the dark fermion $\chi$ and the scalar sector. The dotted regions indicate the exclusion ranges for the dark fermion mass, i.e., $m_\chi<1.3$ GeV and $m_\chi>7.76$ GeV, with corresponding value of $\lambda_D$ that varies within $0.005 - 0.15$, from the sufficient self-interaction dark matter analysis (see Ref. \cite{Dutta:2022knf} for the detailed analysis). Due to this constraint from their analysis, we note that the light scalar mass {$m_{H_3}< 45$ GeV} is strongly disfavored.

\section{Summary and outlook} \label{sec6}
In this work, we have studied the vacuum structure and stability of an extended scalar model featuring a $U(1)_D$ global symmetry. Besides the SM scalar Higgs doublet, the scalar sector contains two a new additional doublets, a real and a complex singlet. This introduces a richer phenomenology and potential vacuum configurations compared to those of the SM. 

We derived the scalar mass spectrum by focusing on the CP-even scalar sectors, providing the expressions for the physical states and their masses. Theoretical constraints, including copositivity conditions and the requirement of a global minimum, were imposed to ensure a stable and bounded potential, resulting in the viable grid-like parameter space shown in Figure \ref{fig:5}. We also numerically examined the perturbative limits of the chosen couplings to satisfy the previous constraints and confirm the validity of the model at various energy scales. This data is also shown in Figure \ref{fig:5}, denoted by a red region.

Our numerical study, based on renormalization group equations, demonstrated that the scalar potential remains stable up to the Planck scale, with the Higgs quartic coupling $\lambda_{SM}\equiv \lambda_4 > 0$ throughout the energy range. This stability is achieved while respecting both theoretical bounds and experimental constraint, i.e., the Higgs invisible decay width. The analysis also revealed viable regions of parameter space where the vacuum structure is consistent with these proposed criteria. In particular, we found that the favored mass region of $m_{H_{3}}$ is within {$45-60$ GeV for the Yukawa coupling $\lambda_D < 0.15$}.

As an outlook of this work, analyzing loop-corrected effective potentials can be involved. In addition, the resulting physically massless pseudoscalar $A_1$ can have theoretical and phenomenological constraints that should be satisfied to ensure consistency with the observations. In view of the theoretical constraint, if the pseudoscalar couples to gauge fields (e.g., gluons or photons), the associated symmetry must not be anomalous unless we aim to address the strong CP problem. To maintain a consistent theory, anomalies involving Axion-like particles must cancel out. On the phenomenological side, a massless pseudoscalar can mediate a long-range force and affect the particle decay. Precision measurements (e.g., in rare decays and electric dipole moments) impose strong constraints on coupling to gauge fields. These couplings must be suppressed to avoid conflicts with experimental results. Moreover, if the massless pseudoscalar is produced in the early universe and decoupled late (e.g., around or after neutrino decoupling), it poses additional cosmological constraints. As a result, it will contribute to the radiation energy density (measured via $N_{\text{eff}}$) and must not overclose the universe. Such overclosure could disrupt nucleosynthesis or lead to excessive entropy production. A comprehensive study is required to explore such constraints in more detail, and this can be another viable future direction for the investigation.

\vspace{1.0cm}
\noindent {\bf Acknowledgment} \\
Y.K.A. thanks the theoretical high-energy physics research group of the Research Center for Quantum Physics for their generous hospitality during the research activities. {A.S.A. would like to thank Physics and Ideas On the New frontier (PION) workshop 2025 for stimulating discussion, where the last part of this work has been completed.} This work was supported by and initiated at BRIN through the Research Assistantships Program.

\appendix
\section{BFB conditions} \label{app.A}
In this appendix, we present the bounded-from-below conditions via copositivity for the quartic potential defined in Eq.\eqref{eq.18}. Depending on the signs of the off-diagonal elements, we classify eight different cases of necessary conditions \cite{PING1993109} (but not sufficient) to ensure the stability of the potential at large field values. These conditions read

\begin{center}
    \it Case I
\end{center}
If all the off-diagonal elements are positive, the matrix is copositive if and only if
\begin{equation}
    \l_1, \l_4, \l_6, \l_5\geq0. \tag{A.1}
\end{equation}

\begin{center}
    \it Case II
\end{center}
If $\l_2 + \l_3 + \l_{10} \equiv \l_{231}\leq0$, the matrix is copositive iff
\begin{equation}
    \l_1\l_4 - \l_{231}^2\geq0. \tag{A.2}
\end{equation}

\begin{center}
    \it Case III
\end{center}
If $\l_{231},\frac{\l_9}{2}\leq0$, the matrix is copositive iff
\begin{align}
    \l_1\l_4 - \l_{231}^2 &\geq0, \nonumber \\
    \l_5\l_6-\frac{1}{4}\l_9^2 &\geq0. \tag{A.3}
\end{align}

\begin{center}
    \it Case IV
\end{center}
If $\l_{231},\l_{11}\leq0$, the matrix is copositive iff
\begin{equation}
    \l_1\l_8 - \l_{231}\l_{11}+\sqrt{(\l_1\l_4-\l_{231}^2)(\l_1\l_6-\l_{11}^2)}\geq0. \tag{A.4}
\end{equation}

\begin{center}
    \it Case V
\end{center}
If $\l_{231},\l_8,\l_{11}\leq0$, the coupling matrix is copositive when
\begin{align}
    \l_1,\l_4,\l_6 &\geq 0, \nonumber \\
    \bar{\l}_{12}\equiv\l_{231} +\sqrt{\l_1\l_4} &\geq0, \nonumber \\
    \bar{\l}_{13}\equiv\l_{11}+\sqrt{\l_1\l_6} &\geq0, \nonumber \\
    \bar{\l}_{23}\equiv\l_8+\sqrt{\l_4\l_6} &\geq0, \nonumber \\
    \l_{231}\sqrt{\l_6} + \l_{11}\sqrt{\l_4} + \l_8\sqrt{\l_1} +\sqrt{\l_1\l_4\l_6} + \sqrt{2\bar{\l}_{12}\bar{\l}_{13}\bar{\l}_{23}} &\geq0. \tag{A.5}
\end{align}

\begin{center}
    \it Case VI
\end{center}
If $\l_{231},\l_8,\l_{12}\leq0$, the coupling matrix is copositive iff
\begin{align}
    \l_1\l_4-\l_{231}^2 &\geq 0, \nonumber \\
    \l_1\l_6-\l_{11}^2 &\geq 0, \nonumber \\
    \l_1\l_5-\l_{12}^2 &\geq 0, \nonumber \\
    \bar{\l}_{12}\equiv \l_1\l_8 - \l_{231}\l_{11} + \sqrt{(\l_1\l_4-\l_{231}^2)(\l_1\l_6-\l_{11}^2)} &\geq0,  \nonumber\\
    \bar{\l}_{13}\equiv \frac{1}{2}\l_1\l_7 - \l_{231}\l_{12} + \sqrt{(\l_1\l_4-\l_{231}^2)(\l_1\l_5-\l_{12}^2)} &\geq0, \nonumber\\
    \bar{\l}_{23}\equiv \frac{1}{2}\l_1\l_9 - \l_{11}\l_{12} + \sqrt{(\l_1\l_6-\l_{11}^2)(\l_1\l_5-\l_{12}^2)} &\geq0, \nonumber\\
    (\l_1\l_8-\l_{231}\l_{11})\sqrt{\l_1\l_5-\l_{12}^2} + \left(\frac{1}{2}\l_1\l_7 - \l_{231}\l_{12}\right) & \nonumber\\
    \sqrt{\l_1\l_6-\l_{11}^2}+\left(\frac{1}{2}\l_1\l_9-\l_{11}\l_{12} \right)\sqrt{\l_1\l_4-\l_{231}^2} + \sqrt{2\bar{\l}_{12}\bar{\l}_{13}\bar{\l}_{23}} & \nonumber\\
    + \sqrt{(\l_1\l_4-\l_{231}^2)(\l_1\l_6-\l_{11}^2)(\l_1\l_5-\l_{12}^2)} &\geq0. \tag{A.6}
\end{align}

\begin{center}
    \it Case VII
\end{center}
If $\l_{231},\l_8,\frac{1}{2}\l_9\leq0$,
    \begin{align*}
        \l_6\left(\l_4\l_{11}^2-2\l_{231}\l_{11}\l_8+\l_1\l_8^2\right) &\geq 0, \nonumber\\
        \l_4\l_6-\l_8^2 &\geq 0, \nonumber\\
        \l_5\l_6 - \frac{1}{4}\l_9^2 &\geq 0, \nonumber\\
        \bar{\l}_{12}\equiv \l_6\left(\l_4\l_{11}-\l_{231}\l_8\right) &\nonumber\\
        + \sqrt{\l_6\left(\l_4\l^2_{11}-2\l_{231}\l_8\l_{11}+\l_1\l_8^2\right)(\l_4\l_6-\l_8^2)} &\geq 0,\nonumber \\
        \bar{\l}_{13}\equiv\l_6\left(\frac{1}{2}\l_7\l_{11}-\l_8\l_{12}\right) &\nonumber\\
        + \sqrt{\l_6\left(\l_4\l^2_{11}-2\l_{231}\l_8\l_{11}+\l_1\l_8^2\right)\left(\l_4\l_6-\frac{1}{4}\l_9^2\right)} &\geq 0, \nonumber\\
        \bar{\l}_{23}\equiv\frac{1}{2}(\l_6\l_7-\l_8\l_9) + \sqrt{\left(\l_4\l_6-\l_8^2\right)\left(\l_5\l_6-\frac{1}{4}\l_9^2\right)} &\geq 0, \nonumber\\
        \l_6\left(\l_4\l_{11}-2\l_{231}\l_8\right)\sqrt{\l_5\l_6-\frac{1}{4}\l_9^2} + \l_6\left(\frac{1}{2}\l_7\l_{11}-\l_8\l_{12}\right)\sqrt{\l_4\l_6-\l_8^2} & \nonumber\\
        +\frac{1}{2}(\l_6\l_7-\l_8\l_9)\sqrt{\l_6\left(\l_4\l^2_{11}-2\l_{231}\l_8\l_{11}+\l_1\l_8^2\right)} &\nonumber\\
        \sqrt{\l_6\left(\l_4\l^2_{11}-2\l_{231}\l_8\l_{11}+\l_1\l_8^2\right)(\l_4\l_6-\l_8^2)\left(\l_5\l_6-\frac{1}{4}\l_9^2\right)}+\sqrt{2\bar{\l}_{12}\bar{\l}_{13}\bar{\l}_{23}} &\geq 0. \tag{A.7}
    \end{align*}

\begin{center}
    \it Case VIII
\end{center}
If $\l_{231},\l_8,\frac{1}{2}\l_9,\l_{12}\leq0$,
    \begin{align*}
        \l_5\left(\frac{1}{4}\l_1\l_{7}^2-2\l_{231}\l_12\l_7+\l_4\l_{12}^2\right) &\geq 0, \nonumber\\
        \l_1\l_5-\l_{12}^2 &\geq 0, \nonumber\\
        \l_5\l_6 - \frac{1}{4}\l_9^2 &\geq 0, \nonumber\\
        \bar{\l}_{12}\equiv \l_5\left(\frac{1}{4}\l_1\l_{7}-2\l_{231}\l_{12}\right) & \nonumber\\
        +\sqrt{\l_5\left(\frac{1}{4}\l_1\l_{7}^2-2\l_{231}\l_{12}\l_7+\l_4\l_{12}^2\right)(\l_1\l_5-\l_{12}^2)} &\geq0, \nonumber\\
        \bar{\l}_{13}\equiv \l_5\left(\frac{1}{2}\l_7\l_{11}-\l_8\l_{12}\right) &\nonumber\\
        +\sqrt{\l_5\left(\frac{1}{4}\l_1\l_{7}^2-2\l_{231}\l_{12}\l_7+\l_4\l_{12}^2\right)\left(\l_5\l_6-\frac{1}{4}\l_9^2\right)} &\geq0, \nonumber\\
        \bar{\l}_{23}\equiv \l_5\l_{11}-\frac{1}{2}\l_9\l_{12}+\sqrt{(\l_1\l_5-\l_{12}^2)\left(\l_5\l_6-\frac{1}{4}\l_9^2\right)} &\geq0, \nonumber\nonumber\\
        \l_5\left(\frac{1}{4}\l_1\l_{7}-2\l_{231}\l_{12}\right)\sqrt{\l_5\l_6-\frac{1}{4}\l_9^2}+\l_5\left(\frac{1}{2}\l_7\l_{11}-\l_8\l_{12}\right)\sqrt{\l_1\l_5-\l_{12}^2} &\nonumber\\
        +\left(\l_5\l_{11}-\frac{1}{2}\l_9\l_{12}\right)\sqrt{\l_5\left(\frac{1}{4}\l_1\l_{7}^2-2\l_{231}\l_{12}\l_7+\l_4\l_{12}^2\right)} &\nonumber\\
        +\sqrt{\l_5\left(\frac{1}{4}\l_1\l_{7}^2-2\l_{231}\l_{12}\l_7+\l_4\l_{12}^2\right)\left(\l_5\l_6-\frac{1}{4}\l_9^2\right)} + \sqrt{2\bar{\l}_{12}\bar{\l}_{13}\bar{\l}_{23}} &\geq0. \tag{A.8}
    \end{align*}

\section{One-loop RG equations} \label{app.B}
Below we provide the one-loop RG equations for the remaining gauge $(g_1, g_2, g_3)$ and Yukawa $(y_t, y_{\nu_R}, \lambda_D, \lambda_\chi)$ couplings involved in the present setup. These RGEs are derived using the PyR@TE 3 package \cite{Sartore:2020gou}. 
Since the new scalars do not carry any color charges, no modification is required in the RG function of the SM gauge coupling $g_3$ of the strong interaction.
The RG equations of the gauge couplings are given by,
\begin{align}
    16\pi^2 \frac{dg_1}{d\mu} &= \frac{43}{6} g_1^3, \\
    16\pi^2 \frac{dg_2}{d\mu} &= -\frac{17}{6} g_2^3, \\
    16\pi^2 \frac{dg_3}{d\mu} &= - 7 g_3^3,
\end{align}
while the RG equations of the Yukawa couplings are
\begin{align}
    16\pi^2 \frac{dy_t}{d\mu} &= \frac{3}{2} y_t^3 - \frac{17}{12} g_1^2 y_t -\frac{9}{4} g_2^2 y_t - 8 g_3^2 y_t,\\
    16\pi^2 \frac{dy_{\nu_R}}{d\mu} &= \frac{3}{2} y_{\nu_R}^3 + \frac{1}{2} \lambda_\chi^2  y_{\nu_R} + y_{\nu_R}^3 - \frac{3}{4} g_1^2  y_{\nu_R} - \frac{3}{4} g_2^2 y_{\nu_R}, \\
    16\pi^2 \frac{d\lambda_D}{d\mu} &= 32 \lambda_D^3 + 4 \lambda_D \lambda_\chi^2, \\
    16\pi^2 \frac{d\lambda_\chi}{d\mu} &= 2  \lambda_D^2  \lambda_\chi + \frac{1}{2}  \lambda_\chi  y_{\nu_R}^2 + \frac{3}{2}  \lambda_\chi^3 + \lambda_\chi^3 - \frac{3}{4} g_1^2 \lambda_\chi - \frac{9}{4}  g_2^2 \lambda_\chi.
\end{align}



\end{document}